\documentclass[12pt]{iopart}

\usepackage{graphicx}
\usepackage{psfrag}
\usepackage{epstopdf}
\begin{document}

\title{Zero-Range Processes with Multiple Condensates:
Statics and Dynamics}
\author{Y. Schwarzkopf$^{1,2}$, M. R. Evans$^3$ and  D. Mukamel$^{1}$}

\address{
$^1$Department of Physics of Complex Systems, Weizmann Institute of Science,
\\ Rehovot, Israel 76100.\\
$^2$Present address: Department of Physics, California Institute of Technology, Pasadena, California 91125.\\
$^3$SUPA, School of Physics, University of Edinburgh, \\
Mayfield Road, Edinburgh EH9 3JZ, UK.}
\ead{yoni@caltech.edu, m.evans@ed.ac.uk,
david.mukamel@weizmann.ac.il}

\begin{abstract}
The steady-state distributions and dynamical behaviour of Zero Range
Processes with hopping rates which are non-monotonic functions of the
site occupation are studied.  We consider two classes of non-monotonic
hopping rates. The first results in a condensed phase containing a
large (but subextensive) number of mesocondensates each containing a
subextensive number of particles.  The second results in a condensed
phase containing a finite number of extensive condensates.  We study
the scaling behaviour of the peak in the distribution function
corresponding to the condensates in both cases. In studying the
dynamics of the condensate we identify two timescales: one for
creation, the other for evaporation of condensates at a given site. The
scaling behaviour of these timescales is studied within the Arrhenius
law approach and by numerical simulations.

\end{abstract}


\date{\today}
\maketitle
\section{Introduction}
Real-space condensation is a phenomenon which occurs in a variety of
physical systems such as jamming in traffic flow, granular clustering,
wealth condensation and hub formation in complex networks \cite{BBJ97,Evans00}.
In a condensation process a finite fraction of the microscopic
constituents aggregate in space.  A simple (minimal) model which has
commonly been used to describe this phenomenon in recent years, is the
Zero-Range Process (ZRP)\cite{Spitzer}.  The model comprises particles
distributed over the sites of a lattice, with stochastic dynamics allowing
particles to hop between sites. The hopping rate is just a function of
the occupation, $n$, of the site from which a hop takes place, hence the name
ZRP. For  recent reviews of the  properties  of the ZRP
see \cite{EH05,Godreche06}.

An appealing feature of the ZRP is that its steady-state distribution
is exactly calculable in terms of the hopping rates. It has been shown
that for a class of hopping rates which are monotonic and decrease
suitably slowly with $n$, the model exhibits a condensation transition
at some critical value of the particle density (the average number of
particles per lattice site). Below the critical density the system is
in the fluid phase where all sites are equivalent and have a
characteristic occupation given by the average density. Above the
critical density the excess particles---a finite fraction of the total
number of particles---condense onto a {\em single} site. The
supercritical system is thus composed of a critical fluid coexisting
with a condensate.

A choice of hopping rate commonly used to study condensation is
\begin{eqnarray}\label{ucond}
u(n) &=& 1 + \frac{b}{n}\quad\mbox{for}\quad n>0\;,\\
u(0) &=&  0 \;.\nonumber
\end{eqnarray}
For $b<2$ it has been shown that there is no condensation at any density,
while for $b>2$ condensation takes place above a critical density
\cite{EH05,Godreche06,GSS03,Godreche03}
\begin{equation}
\rho_c = \frac{1}{b-2}\;.
\end{equation}

Our interest in this work is to consider dynamics which could lead to
the formation of more than one condensate.  Previous studies have
shown that multiple condensates can exist in models with
non-conserving dynamics where in addition to the above hopping rate
(\ref{ucond}), local creation and annihilation processes are
introduced \cite{AELM05,AELM07}.
Here we consider ZRP models with {\em conserving}
dynamics (i.e. where particles are neither created nor destroyed)
which generate multiple condensates.  A simple mechanism for this is to choose
hopping rates which are no longer monotonic but increase for large
$n$. This effectively introduces a soft cut-off in the
occupation number of a site.  Then, the excess particles are forced to
be shared over multiple condensate sites.  We study both steady-state
properties and the dynamical features corresponding to creation and
evaporation of condensates in such a model.

The paper is organised as follows. The model and different dynamics we
consider are defined and general steady state properties are reviewed
in Section~\ref{moddef}. In Section~{\ref{secalg} we analyse the
properties of the  steady state in the case of an algebraically
increasing cut-off in the hop rates.  In Section~{\ref{secexp} we
analyse the properties of the  steady state in the case of an
exponentially increasing cut-off in the hop rates.  In particular the
scaling of the condensate properties are compared with numerical
simulations. The dynamics of the condensates are studied in
Section~\ref{dynamics}. Finally, some conclusions are drawn in
Section~\ref{conc}.
%%%%%%%%%%%%%%%%%%%%%%%%%%%%%%%%%%%
\section{Model definition}
%%%%%%%%%%%%%%%%%%%%%%%%%%%%%%%%%%%
\label{moddef}
We consider a lattice of $L$ sites
with a total of $N$ particles and average density $\rho = N/L$.
Each occupation of a lattice site $l$ is given by $n_l$.
One choice of hopping rate we consider is
given by
\begin{equation}
u(n) =  1 + \frac{b}{n} + c\left( \frac{n}{L}\right)^k
\quad\mbox{for}\quad n>0\;.
\label{u1}
\end{equation}
With this choice, illustrated in Figure~\ref{ufig}, the hopping rate
from a site containing $n= O(L)$ particles is enhanced over the case
(\ref{ucond}). The hopping rate is non-monotonic: initially it
decreases as $1/n$, it has a minimum at $n \sim L^{k/k+1}$ and it
then increases algebraically. We refer to the choice (\ref{u1}) as
algebraic hopping rates. We note that the hopping rates (\ref{u1}) depend on the system size;
another ZRP with size-dependent hopping rates that induce condensation has
been studied in \cite{GS08}.

For $b<2$,  where condensation does not occur, the additional term in
(\ref{u1}) compared to (\ref{ucond}) does not affect the distribution
of occupations.
On the other hand
for $b>2$,  we expect condensation to take place
at the same critical density as for hopping rates (\ref{u1}).
The reason is that the last term in (\ref{u1}) is negligibly small for
nonextensive occupations, therefore does not affect the fluid phase.
Hence, the condensation
transition will be at the same critical density $\rho_c$.
However the  last term in (\ref{u1}) does affect the
condensed phase since the last term becomes significant for
macroscopic $n$ and thus would change the occupation of any
condensate.  We shall show
that, in fact,  in the  condensed phase the system
exhibits  a large number of  mesocondensates whose occupation
increases sublinearly with $L$. The number of mesocondensates
is also sublinear in $L$ but the total occupation
of  all condensates is extensive, accounting for the total number of
excess particles.

Another choice of  hopping rates that we consider  is
\begin{equation}
u(n) =  1 + \frac{b}{n} + \exp(n-aL)
\quad\mbox{for}\quad n>0\;,
\label{u2}
\end{equation}
where $a>0$.
The hopping rate is again
non-monotonic as illustrated in Figure~\ref{ufig}: initially it decreases as $1/n$, it has a minimum at $n \sim aL$ and it
then  increases exponentially.
We refer to the choice (\ref{u2}) as exponential hopping rates.
Here the increase in the hopping rate with $n$ at $n= aL$ is sharper than
(\ref{u1}). In this case the effective cut-off is at an extensive occupation
and, when present, condensates contain up to $aL$  particles.
Thus, at sufficiently high density a finite
number of  condensates each with occupancy $< aL$ may be present.

%%%%%%%%
\begin{figure}[htb]
\begin{center}
\includegraphics[width=10cm]{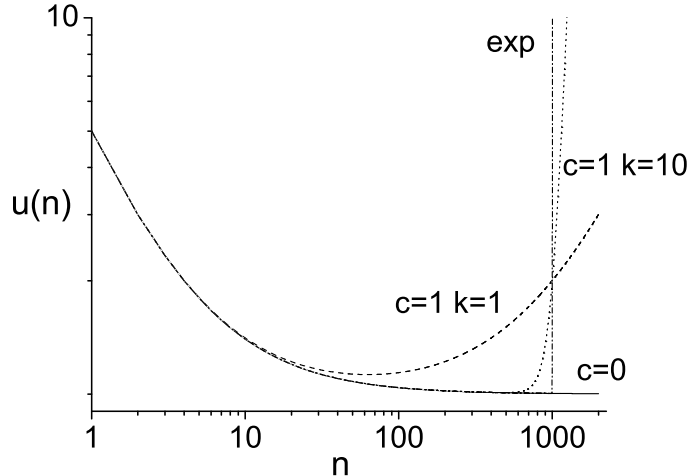}
\caption{\label{ufig} The algebraic hopping rate $u(n)$ given by
(\ref{u1}) %,\ref{u2})
for $b=4$, $L=1000$ and different values of
$k$ and $c$. The full line represents $c=0$ where $u(n)$
is a monotonically decreasing function of $n$ whereas the dashed and
dotted lines represent $c=1$ and $k=1,10$ respectively
where $u(n)$ has a minimum. The exponential hopping rate given by (\ref{u2})
is represented by the dashed-dotted line .
}
\end{center}
\end{figure}
%%%%%%%%

In a numerical simulation of the dynamics, the hopping rates are
implemented in a random sequential updating scheme by first
selecting a site at random. Then one should generate a random number
uniformly distributed between 0 and the maximum possible hopping
rate and carry out a move if the random number is less than or equal
to the rate. If the maximal rate is $u(N)$, this rate   may not be
realised in practice since it would  correspond to all particles
residing in a single site. Therefore in order to speed up the
simulation we choose  a cut-off in the hopping rate so that the
dynamics is not affected.

When a hopping process is carried out the destination site must be
chosen.  In this work we restrict ourselves to one-dimensional totally
asymmetric hopping where the sites are arranged on a ring and
particles hop from site $i$ to the next site $i+1$.  In general, the
steady-state distribution for the ZRP does not depend on the
connectivity although the dynamical properties may do.

%%%%%%%%%%%%%%%%%%%%%%%%%%%%%%%%%%%
\subsection{Steady-State Properties}\label{stst}
%%%%%%%%%%%%%%%%%%%%%%%%%%%%%%%%%%%
%%%%%%%%
\begin{figure}[htb]
\begin{center}
\includegraphics[width=10cm]{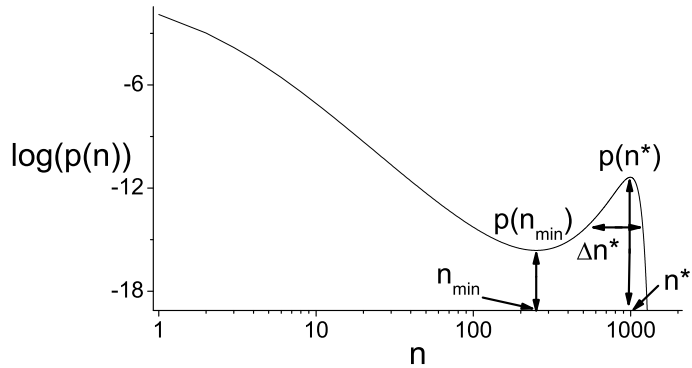}
\caption{ \label{pdetail}A logarithmic plot of 
the single site distribution $p(n)$ for  $L=1600$, $N=6400$
calculated from the exact expression (\ref{pnLN}). The hopping rate
$u(n)$ is  given by the algebraic case (\ref{u1}) with $b=4$,  $c=1$ and $k=10$.}
%fig 13 YS thesis
\end{center}
\end{figure}
%%%%%%%%

The steady state of a ZRP is known exactly
and has a factorised form:
\begin{equation}
P(n_1,\cdots,n_L) = Z^{-1}(L,N)
\prod_{i=1}^{L}f(n_i) \delta(\sum_{j=1}^L n_j- N)
\end{equation}
where  the normalisation $Z(L,N)$, equivalent to a canonical partition function
for $N$ particles and $L$ sites, is defined as
\begin{equation}
Z(L,N) = \prod_{i=1}^{L}\left[ \sum_{n_i=0}^N
f(n_i)\right] \delta(\sum_{j=1}^L n_j- N)\;,
\end{equation}
where the delta symbol ensures that only configurations with precisely
$N$ particles contribute.
The single site distribution,
which is  the probability that
in the steady state a given site contains $n$ particles,
is given by
\begin{equation}
p(n) = f(n) \frac{Z(L-1,N-n)}{Z(L,N)}\;.
\label{pnLN}
\end{equation}
To  compute $Z(L,N)$ and $p(n)$ exactly for small systems
the following recursion is useful
\begin{equation}
Z(L,N)= \sum_{n=0}^N f(n) Z(L-1,N-n)\;.
\label{recur}
\end{equation}
We will use this formulation (namely, the Canonical Ensemble
formulation in which the total  number of particles is fixed) to
provide numerical results. This data will be  compared in
Sections~\ref{numsim} and~\ref{numsim_Exp} with the analytical
predictions of Sections~\ref{gca:alg} and \ref{gca:exp} that are made within the grand canonical ensemble which we
now describe.

To describe analytically the properties of $p(n)$ it is convenient
to work  within the grand canonical ensemble (where the total number
of particles is allowed to fluctuate and one introduces a fugacity $z$
to control the number of particles). The steady-state distribution
becomes
\begin{equation}
P(n_1,\cdots,n_L) = \prod_{l=1}^L p(n_l)\;,
\end{equation}
where
\begin{equation}
p(n) = A z^n f(n)\;,
\label{pn}
\end{equation}
and
\begin{equation}
f(n) = \prod_{m=1}^n \frac{1}{u(m)}\;.
\label{fn}
\end{equation}
The normalization constant $A$ in (\ref{pn}) is chosen to ensure that
sum of $p(n)$ is equal to 1.  The fugacity $z$ is determined by the
condition that the mean total number of particles in the system is $N$.

In the commonly studied case with $u(n)$ given by
(\ref{ucond}), the probability distribution at large $n$
becomes
\begin{equation}
p(n) \simeq A \frac{z^n}{n^b}\;.
\end{equation}
The condensation transition is easy to understand.
For densities below the critical density, $\rho_c$, the fugacity
satisfies $z<1$ and the probability decays exponentially for large
$n$. This distribution describes what is termed the fluid phase.  As
the critical density is approached, $z$ approaches 1 and the decay of
$p(n)$ becomes algebraic, corresponding to a critical fluid.  Above
the critical density an extra piece of $p(n)$  emerges representing the
condensate \cite{MEZ05,EMZ06}.  Therefore the condensate co-exists
with the critical fluid which has density $\rho_c$.  The number of
particles in the condensate is given by $L(\rho-\rho_c)$.

%%%%%%%%%%%%%%%%%%%%%%%%%%%%%%%%%%%
\section{Analysis of condensation for algebraically increasing $u(n)$}
\label{secalg}
%%%%%%%%%%%%%%%%%%%%%%%%%%%%%%%%%%%
In this Section we study the characteristic features of the single
site occupation distribution $p(n)$ for the case of algebraically
increasing hopping rate (\ref{u1}).   In the
condensed phase we expect this distribution to have the general form
given in Figure~\ref{pdetail}. The distribution decays algebraically
for intermediate $n$ and there is a peak in the distribution for
large $n$.  The peak is associated with any condensates in the
system. Our aim is to analyse the shape of this peak---its height,
width and position.  This information will determine the typical
number of condensates present in the system and their typical
occupancy.

\subsection{Grand Canonical Analysis}\label{gca:alg}
To investigate  the probability distribution
in the condensed phase
within the grand canonical ensemble we consider the effective
potential $\Phi(n)$ defined through
\begin{equation}
p(n)=A \e^{-\Phi(n)}\;.
\end{equation}
Using
(\ref{pn},\ref{fn}),
$\Phi(n)$  is  given by
\begin{equation}
\Phi(n) =  -n \ln z + \sum_{m=1}^n \ln u(m)\;.
\end{equation}
Expanding $\ln u(m)$ for large $m$  and  approximating for $n \ll L$ the sum
as an integral yields
\begin{eqnarray}
\sum_{m=1}^n \ln u(m)  &\simeq& -b \ln n  +  \frac{c}{k+1} \frac{n^{k+1}}{L^k}
\;.
\end{eqnarray}
Since we are discussing condensation we consider
only hopping rates with $b>2$ for which condensation is possible.  
In the condensed phase we expect the fugacity $z\to 1$ in the
thermodynamic limit. Hence, one describes the fugacity as a function of
the system size $L$, such that
\begin{equation}
z=\e^{h(L)}\;,
\end{equation}
where $ h(L) \to  0$ as $L\to \infty$.

Putting all this together we obtain,  for large $n$ and $L$,
\begin{equation}\label{Phin}
\Phi(n)=-nh(L)\,+\,b\ln n\,+\,\frac{c}{k+1}\frac{n^{k+1}}{L^k}\;,
\end{equation}
which implies \begin{equation}
p(n) \simeq   \frac{A}{n^b}  \exp \left[n h(L)- \frac{c}{k+1} \frac{n^{k+1}}{L^k}
\right]
\end{equation}
where $A$ is a normalization constant.

We now identify  the position of the peak
which corresponds to a maximum of $p(n)$  at $n^*$.
The condition
\begin{equation}
\Phi'(n)=
-h(L)+\frac{b}{n} + c\frac{n^{k}}{L^k}=0\;,
\label{phid0}
\end{equation}
gives
\begin{equation}
h(L) = \frac{b}{n^*} + c \frac{(n^*)^k}{L^k}\;.
\end{equation}
Then  $h(L)$ can be eliminated  in (\ref{Phin}) resulting in
\begin{equation}
\Phi(n^*) = - \frac{ck}{k+1} \frac{(n^*)^{k+1}}{L^k} + b \ln n^* -b \;.
\label{phinst}
\end{equation}
Looking for a solution such that  $p(n^*)$ scales as a power law of
$L$ implies that $\Phi(n^*) = O(\ln L)$. The leading order term in
Equation (\ref{phinst}) is the first term, and therefore
\begin{equation}
 n^*          \sim   L^{k/(k+1)}\,\left[\ln L\right]^{1/(k+1)}\;,
\label{nstar}
\end{equation}
which implies that the leading order of $h(L)$
is
\begin{equation}
h(L) \sim  \left( \frac{\ln L}{L}\right)^{k/k+1}\;.
\end{equation}

The width of the peak at $n^*$ is given by
\begin{equation}
\Delta n^*=\bigg[\Phi''(n^*)\bigg]^{-1/2}\;,
\end{equation}
where
\begin{equation}
\Phi''(n) = ck \frac{n^{k-1}}{L^k} - \frac{b}{n^2}\;.
\end{equation}
Using the above expression for $n^*$, and noting that the first term
of this expression is the leading one for large $L$, we find
\begin{equation}
 \Delta n^*   \sim  k^{-1/2} L^{k/(k+1)}\, \left[\ln L\right]^{-(k-1)/2(k+1)}\;.\label{delnstar}
\end{equation}
Since $\Delta n^*/n^* \sim 1/\left[ \ln L\right]^{1/2}$ vanishes
in the large $L$ limit we expect the canonical and grand canonical ensembles
to be equivalent.

In order to calculate $p(n^*)$ we note that
the weight, $w$, of the peak, that is
the probability contained in the peak, is given by
\begin{equation}
w \sim p(n^*) \Delta n^*\;.
\end{equation}
In the condensed phase,
the weight $w$ represents the fraction of sites which are condensate sites.
The fraction of particles contained within the  condensates
is then $n^*w$.
Since in the condensed phase this fraction is finite (i.e. $\rho-\rho_c$)
we arrive at the condition
\begin{equation}
p(n^*) n^* \Delta n^*  = O(1)
\end{equation}
yielding
\begin{equation}
 p(n^*)       \sim k^{1/2}  L^{-2k/(k+1)} \left[\ln L\right]^{(k-3)/2(k+1)}\;.
\label{pnstar}
\end{equation}
The weight of the peak, $w$ thus scales as
\begin{equation}
w       \sim   L^{-k/(k+1)} \left[\ln L\right]^{(1/(k+1)}\;.
\label{w1}
\end{equation}

The above analysis of the scaling of the peak in $p(n)$ implies the
following condensation behaviour.  We first note that the size of the
condensates, (\ref{nstar}), scales sub-linearly with the system size
and therefore these are termed mesocondensates.  The typical number of
mesocondensates is $Lw$ which scales as $L^{1/(k+1)}$ up to
logarithmic corrections.  Therefore their number diverges sublinearly
with $L$.

We conclude by investigating the dip in $p(n)$ to the left of $n^*$
denoted $n_{min}$ in Figure~\ref{pdetail}. As we discuss  in
Section~\ref{dynamics} the dip probability is significant in determining the
dynamics of the mesocondensates.  Balancing the first two terms in the
extremum condition (\ref{phid0}) yields
\begin{equation}
n_{min} \sim \frac{1}{h(L)} \sim L^{k/(k+1)}\left[\ln L\right]^{-k/(k+1)}\;.
\label{nmin}
\end{equation}
Then, $p(n_{min})$ is readily evaluated as
\begin{equation}
p(n_{min}) \sim  L^{-bk/(k+1)}\left[ \ln L\right]^{bk/(k+1)}\;.
\label{pnmin}
\end{equation}

In summary,
the leading behaviour
in the system size $L$ of the shape of peak is given by equations
(\ref{nstar},\ref{delnstar},\ref{pnstar}) and the dip
by (\ref{nmin},\ref{pnmin}).

The analysis of this subsection has been carried out within the
grand canonical ensemble. In the following subsection we will
compare it with exact numerical results for finite systems obtained
in the canonical ensemble. This will provide evidence that at least
the leading behaviour for the scaling properties of the condensate
peak are correctly given by
(\ref{nstar},\ref{delnstar},\ref{pnstar}).

%%%%%%%%%%%%%%%%%%%%%%
\subsection{Numerical results}
\label{numsim}
%%%%%%%%%%%%%%%%%%%%%%
For $u(n)$ given by
(\ref{u1}) and for   $\rho>\rho_c$, we plot
in Figure~\ref{pfig}  the exact $p(n)$ for
$L=1000$. This was computed by iterating the recursion relation (\ref{recur})
and using expression (\ref{pnLN}).
%%%%%%%%
\begin{figure}[htb]
\begin{center}
\includegraphics[width=10cm]{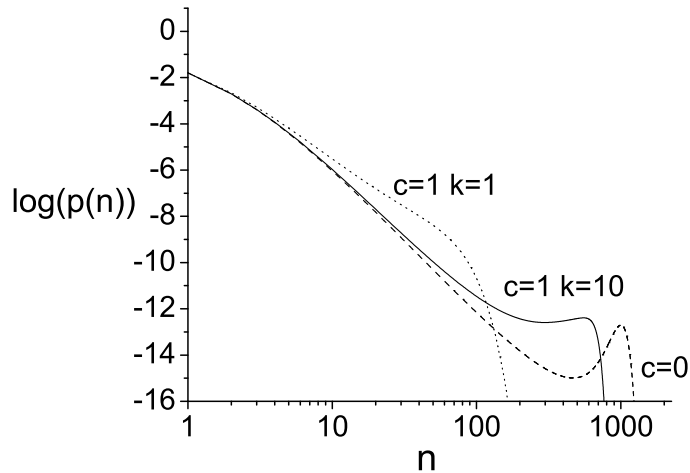}
\caption{\label{pfig} A logarithmic plot of $p(n)$ for  $L=1000$
calculated from the exact expression (\ref{pnLN}). The hopping rate
$u(n)$ is  given by the algebraic case  (\ref{u1}) with
$b=3$, $\rho=2$  and various  values
of $c$ and $k$. The dashed line corresponds to $c=0$, the dotted
line to $c=1,k=1$ and the full line to $c=1,k=10$%Fig 9 YS thesis
}
\end{center}
\end{figure}
%%%%%%%%
As expected the probability distribution exhibits a power law decay
representing the fluid and a peak corresponding to the condensates.
In the $c=0$ case, there is a single condensate containing
$L[\rho-\rho_c]$ particles which results in a peak location that
scales as $n^*\sim L$. For $b>3$, the second moment is finite and the
peak has a width of $\Delta n^* \sim L^{1/2}$.  For $2<b<3$, the regime
of anomalous fluctuations, the second moment diverges and the width of
the peak scales as $\Delta n^* \sim L^{1/(b-1)}$.  For both regimes
the peak is sharp in the sense that $\Delta n^*/n^* \to 0$.  It is
clearly seen in Figure~\ref{pfig} that the peak for the case $c=0$ is
sharper and at larger $n$ than for $c>0$ as expected from the above
and (\ref{nstar},\ref{delnstar},\ref{pnstar}). 

In Figure~\ref{sim} we present some snapshots of the occupations of
the sites in the steady state as obtained from numerical simulations.
It can be seen that while for $c=0$
(where (\ref{u1}) reduces to (\ref{ucond}))
there is a single condensate, for
$c>0$ the excess density above the critical value is distributed among
a number of mesocondensates.
It is the distribution of the occupations of these mesocondensates
that the peak in $p(n)$ describes.

%%%%%%%%
\begin{figure}[htb]
\begin{center}
\includegraphics[width=12cm]{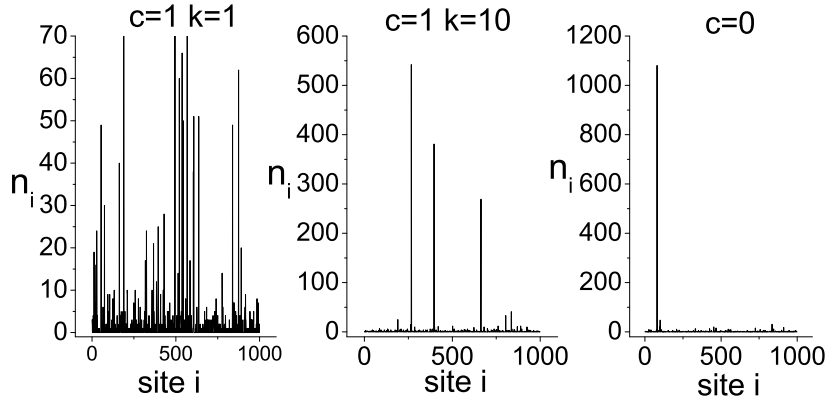}
\caption{\label{sim} Site occupations in snapshots of the system
obtained from numerical
simulation for $L=1000$,  $\rho=2$, in the case of algebraic hop
rates with $b=3$ and several values of $c$
and $k$:  from left to right for $c=1,k=1$, $c=1,k=10$ and
$c=0$.%Fig 10 YS thesis
}
\end{center}
\end{figure}
%%%%%%%%

%%%%%%%%
\begin{figure}[htb]
\begin{center}
\includegraphics[width=10cm]{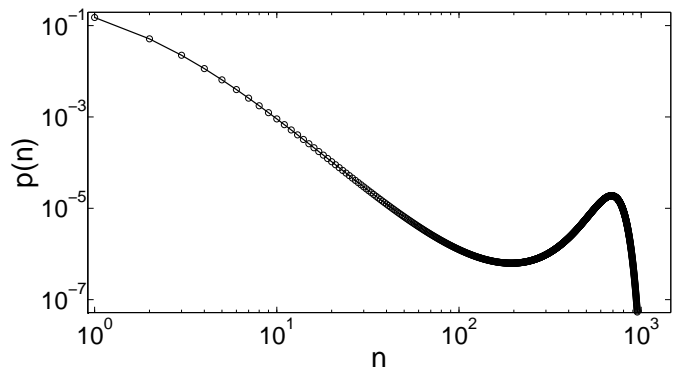}
\caption{\label{grand_alg} A comparison of the canonical probability
(\ref{pnLN}) and the grand canonical probability (\ref{pn}),
 represented by the full line and ($\circ$) respectively.
The distributions were calculated numerically for  algebraic hopping (\ref{u1}) rates
with $L=1600$,  $b=4$, $\rho = 4$
$c=1$ and   $k= 5$. }
\end{center}
\end{figure}
%%%%%%%%

Since the steady state properties of the distribution given in
Section~\ref{secalg} were calculated in the grand canonical
ensemble,we wish to compare the distribution obtained in the canonical ensemble
with the one obtained in the grand canonical ensemble.  In
Figure~\ref{grand_alg} the probability distribution for the parameters
$L=1600$, $b=4$, $\rho=4$, $c=1$ and $k=5$ was calculated numerically both in the
grand-canonical ensemble (\ref{pn}) and in the canonical ensemble
(\ref{pnLN}).  It can be seen
in Figure~\ref{grand_alg} that the distributions obtained form
the two ensembles agree.

%%%%%%%%
\begin{figure}[htb]
\begin{center}
\includegraphics[width=10cm]{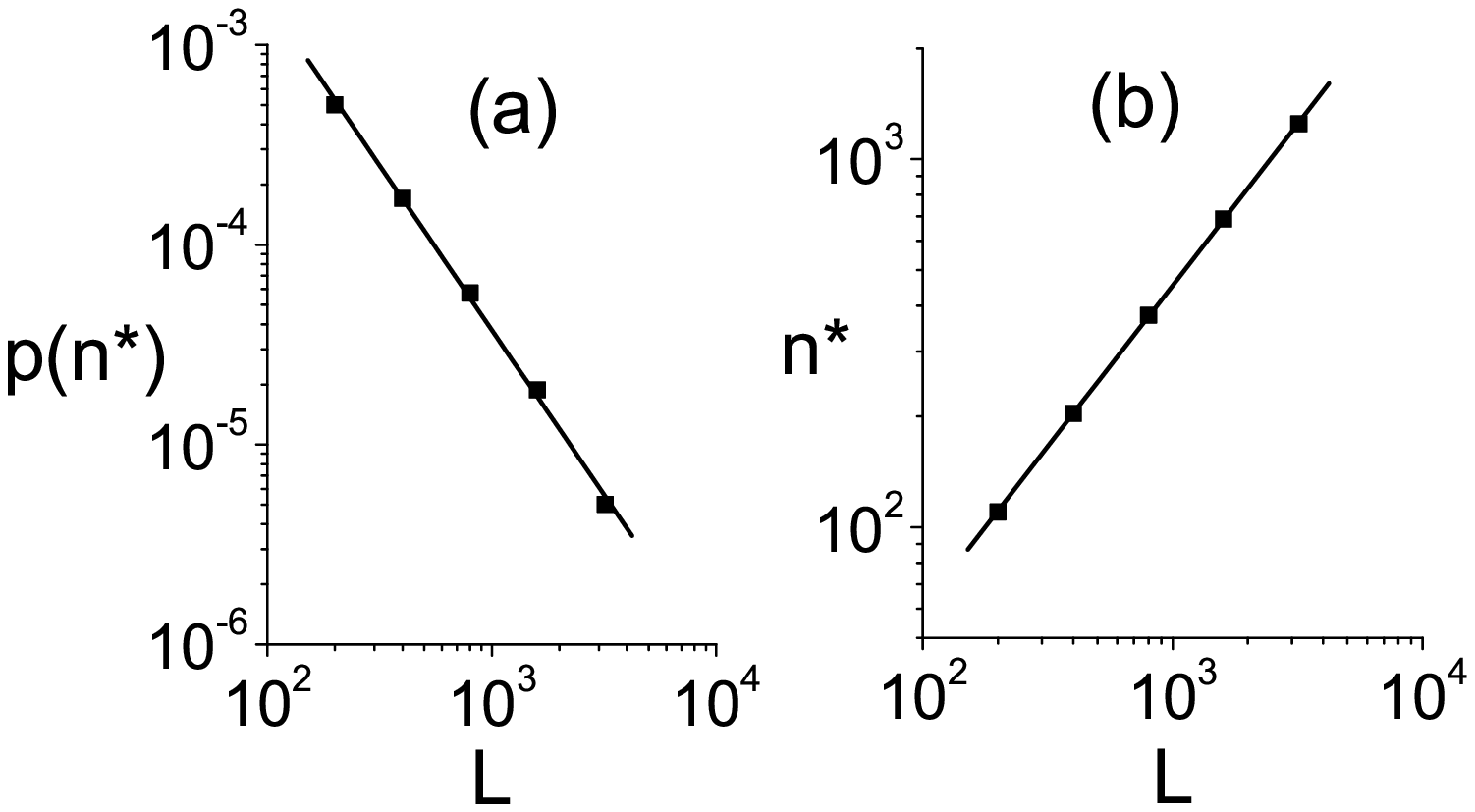}
\caption{\label{scale_nstar_u1}The scaling of the condensate peak as a
function of the system size $L$ in the case of algebraic hopping rates
with $b=4$, $\rho=4$ and $c=1,k=5$.  The probability $p(n^*)$ and
location of the dip $n^*$ were calculated numerically for each system
size $L$ using (\ref{pnLN}) for the hopping rate (\ref{u1}). In (a)
and (b) $p(n^*)$ and $n^*$ are plotted respectively as a a function of
$L$ and are compared to a linear fit of the data points. The data
points for the various system sizes are compared to linear fit with a
slope of $-1.65$ and $0.87$ in (a) and (b) respectively.}

\end{center}
\end{figure}
%%%%%%%%

In Figure~\ref{scale_nstar_u1} we check that the grand canonical
analysis yields the leading scaling properties of the peak calculated
for the canonical distribution in finite size systems.  In
Figure~\ref{scale_nstar_u1}(a), $p(n^*)$ is plotted against $L$.  A
slope which agrees with the expected value of $-1.65 \approx
-2k/(k+1)$ is seen, confirming the $L^{-2k/(k+1)}$ dependence
(\ref{pnstar}).  In Figure~\ref{scale_nstar_u1}(b) $n^*$ is plotted
against $L$.  A slope which agrees with the expected value of $0.87
\approx k/(k+1)$ is seen, confirming the $L^{k/(k+1)}$ dependence
(\ref{nstar}).  The logarithmic corrections were not taken into
 account since the correction $\left[ \ln L \right]^{1/6}$ has negligible
$L$ dependence.

%%%%%%%%
\begin{figure}[htb]
\begin{center}
\includegraphics[width=10cm]{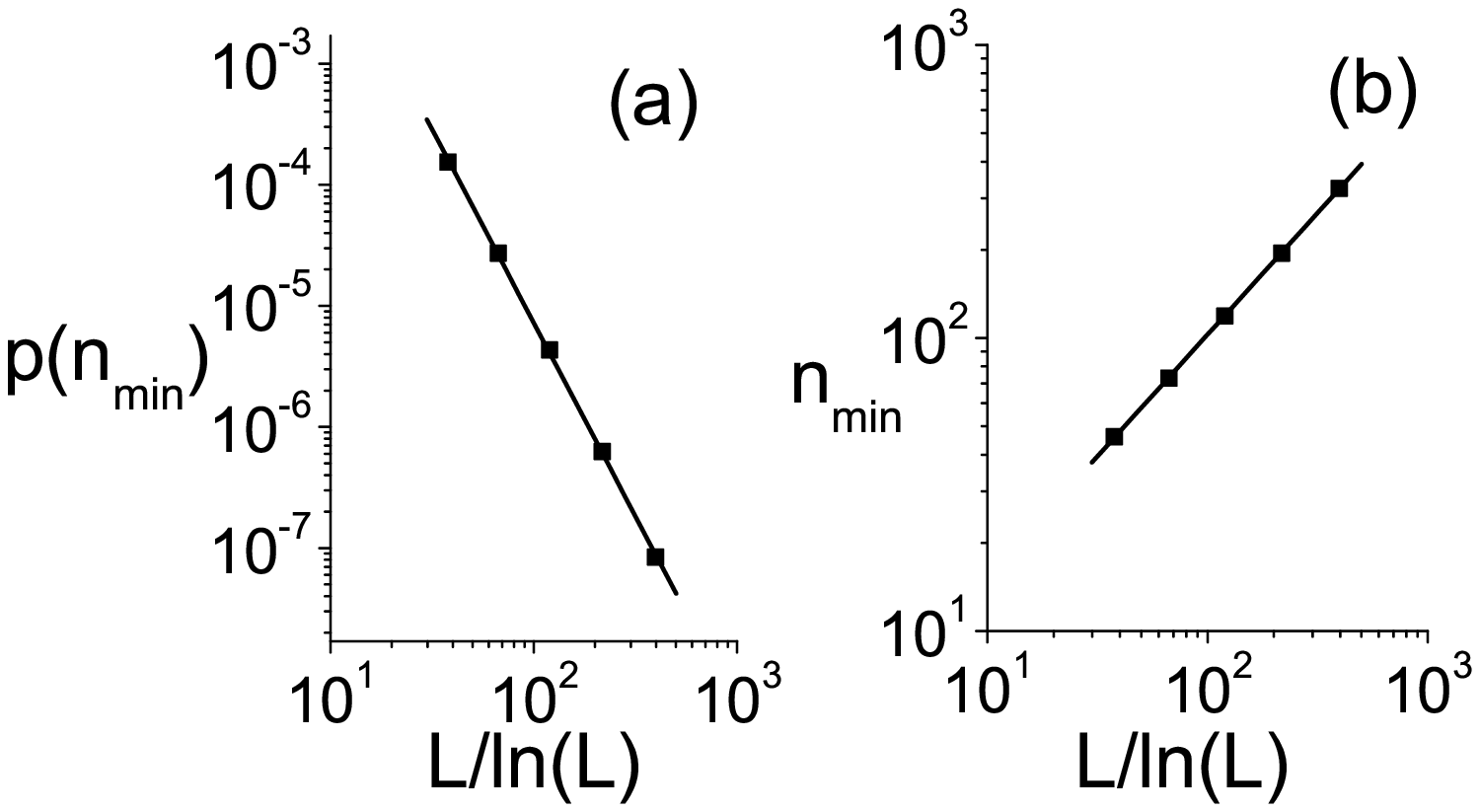}
\caption{\label{scale_nmin} The scaling of the dip as a function of
the system size $L$ in the case of algebraic hopping rates with $b=4$,
$\rho=4$ and $c=1,k=5$.  The probability $p(n_{min})$ and location of
the dip $n_{min}$ were calculated numerically for each system size $L$
using (\ref{pnLN}) for the hopping rate (\ref{u1}). In (a) and (b)
$p(n_{min})$ and $n_{min}$ are plotted respectively as a a function of
$L/\ln(L)$ and are compared to a linear fit of the data points. The
data points for the various system sizes are compare to linear slopes
of $-3.2$ and $0.83$ in (a) and (b) respectively.}
\end{center}
\end{figure}
%%%%%%%%

In Figure~\ref{scale_nmin} we check that the grand canonical analysis yields the leading
scaling properties of the dip calculated for the canonical
distribution in finite size systems. In Figure~\ref{scale_nmin}(a) the dip height
$p(n_{min})$ is plotted against $L/\ln(L)$.  A linear fit for the data
with a slope of $-3.2 \approx -bk/(k+1)$ is seen, 
confirming the $\left[L/\ln(L)\right]^{-bk/(k+1)}$ dependence (\ref{pnmin}).
In Figure~\ref{scale_nmin}(b)
the dip location $n_{min}$ is plotted against $L/\ln(L)$.  A linear fit for
the data with a slope of $0.83 \approx k/(k+1)$  is seen,
 confirming the $\left[L/\ln(L)\right]^{k/(k+1)}$ dependence (\ref{nmin}).

We conclude that  the
steady state properties of the distribution given in
Section~\ref{secalg} computed for the grand canonical ensemble are
able to describe the properties in the canonical ensemble.

%%%%%%%%%%%%%%%%%%%%%%%%%%%%%%%%%%%
\section{Analysis of condensation for 
exponentially increasing $u(n)$}\label{secexp}
%%%%%%%%%%%%%%%%%%%%%%%%%%%%%%%%%%%
In this section we consider  the distribution $p(n)$
corresponding to  hopping rates (\ref{u2}).
As can be seen in Figure~\ref{ufig}, the hopping rate $u(n)$
is basically given by the commonly used form (\ref{ucond}) but with a rapid
increase at $n= aL$. In turn this generates a sharp cut-off in $p(n)$
at $n=aL$.

In the absence of the cut-off the condensate peak is at $n=
(\rho-\rho_c)L$.  Thus, we expect this still to be the case as long
as $(\rho-\rho_c)< a$.  However,  for  densities  above $\rho_c +
a$, and  below $\rho_c + 2a$, the density excess $\rho-\rho_c$ is
distributed among two condensates each containing a maximum of $aL$
particles. In general for densities $\rho_c + q a < \rho \le \rho_c
+ (q+1)a$, where $q$ is  an integer, we would expect to have $q+1$
condensates.
In other words the number of condensates is
\begin{equation}
q + 1 = \lceil \frac{\rho- \rho_c}{a} \rceil\;,
\end{equation}
where $\lceil x \rceil$ denotes the ceiling function which is the
integer part of $x$ plus one.  Thus, in contrast to the algebraic case, this
model exhibits a finite number of extensive condensates.
%%%%%%%%
\begin{figure}[htb]
\begin{center}
\includegraphics[width=10cm]{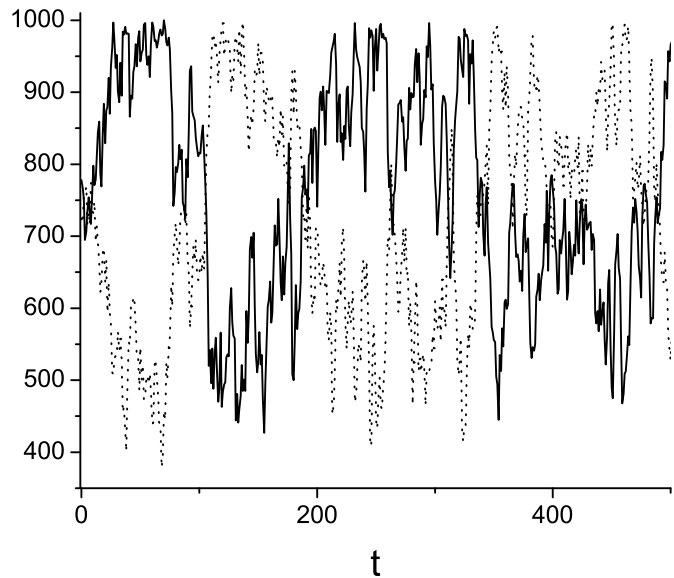}
\caption{\label{cond_oscil} The time-dependent occupation
of the two condensates for exponential hopping rates
with  parameters $L=1000$, $a=1$, $b=4$ and $\rho=2$
calculated from numerical simulation.}
\end{center}
\end{figure}
%%%%%%%%%

For a density $\rho=\rho_c +qa$ with integer $q$, the condensed phase
is composed of $q$ condensates with $aL$ particles each. The structure
of the condensed phase for noninteger $(\rho-\rho_c)/a$ is rather
different.  This is best illustrated by considering the case of two
condensates: $\rho_c + a <\rho < \rho_c + 2a$.  The total number of
particles in the two condensate sites is $(\rho-\rho_c)L$.  Since the
maximum occupation of a site is $aL$, the occupation of a condensate
site fluctuates between $aL$ and $(\rho -\rho_c - a)L$.  A numerical
simulation was carried out for $L=1000$, $a=1$, $b=4$ and $\rho=2$ for
which the expected number of condensates is $\lceil \rho-\rho_c\rceil
= 2$.  In Figure~\ref{cond_oscil} the occupation of the two
condensates is plotted as a function of time.  Interestingly, the two
condensates' occupations fluctuate in an anticorrelated fashion
between $aL$ and $(\rho-\rho_c-a)L$ around the mean occupation
$(\rho-\rho_c)L/2$. As discussed in \cite{JSthesis}, the condensates
interact and the role of the larger condensate is shared among the two
condensates.  This results in a distribution with two peaks with one
centred at $n=aL$ corresponding to $\lfloor (\rho-\rho_c)/a \rfloor$
condensates and the second corresponds to a single condensate
containing the remaining $(\rho-\rho_c-a)L$ particles. Here $\lfloor x
\rfloor$ is the floor function, defined as the highest integer less
than or equal to $x$. The numerical results presented in section
~\ref{numsim_Exp} confirm this behaviour.

In the following subsection we analyze the scaling properties of the
condensates within the grand canonical ensemble. 
The grand canonical analysis of  the ZRP with a sharp cut-off in occupation, as is the case
here,  leads to a peak in the
distribution $p(n)$ at the cut-off for any density above $\rho_c$
\cite{EH05}. On the other hand the considerations of the previous
paragraph indicate that the behaviour of the model with fixed number
of particles (Canonical ensemble) is more complex with a
distribution depending on $\rho$. For example, as pointed out in the
previous paragraph for $\rho-\rho_c <a$ the condensate peak is below
the cut-off.  We conclude that at densities for which $(\rho-
\rho_c)/a$ is non-integer the grand canonical and canonical analyses
do not yield the same scaling behaviour for the condensate. Thus we
apply the grand canonical analysis for the case of {\em integer}
$(\rho- \rho_c)/a$ where the condensate peak is at the cut-off point
$n= aL$. In subsection \ref{numsim_Exp} we numerically check the
equivalence between the two ensembles in this case. 

%%%%%%%%%%%%%%%%%%%%%%%%%%%%%%%%%%%
\subsection{Grand Canonical Analysis}\label{gca:exp}
%%%%%%%%%%%%%%%%%%%%%%%%%%%%%%%%%%%

Following the same
strategy as in subsection \ref{gca:alg}, one finds that for
hopping rate
(\ref{u2})
\begin{equation}\label{Phin2}
\Phi(n)=-nh(L)\,+\,b\ln n\,+ \e^{n-aL}\;,
\end{equation}
which implies
\begin{equation}
p(n) \simeq   \frac{A}{n^b}  \exp \left[n h(L)-  \e^{n-aL}\right]
\end{equation}
where $A$ is a normalization constant.

The extremum condition on $\Phi(n)$ is
\begin{equation}
\Phi'(n)=
-h(L)+\frac{b}{n} + \e^{n-aL}=0\;,
\label{phi2d0}
\end{equation}
which implies
\begin{equation}
\Phi(n^*) = - (n^*-1) \e^{n^*-aL} + b \ln n^* -b \;.
\label{phi2nst}
\end{equation}
Looking for a solution such that   $\Phi(n^*) = O(\ln L)$
implies that
\begin{equation}
 n^*   \simeq  aL - \ln L + \ln \ln L^\alpha
\label{n2star}
\end{equation}
where $\alpha$ is a parameter which in principle can  be determined.
The leading order of $h(L)$
is then
\begin{equation}
h(L) \simeq \e^{n-aL} \simeq \alpha \frac{\ln L}{L}\;.
\end{equation}
To calculate the width of the peak we note that
\begin{equation}
\Phi''(n^*) = -\frac{b}{(n^*)^2} + \e^{n^*-aL} \simeq
\alpha \frac{\ln L}{L}\;,
\end{equation}
yielding
\begin{equation}
\Delta n^* \sim \left[ \frac{L}{\ln L}\right]^{1/2}\;.
\label{delnstar2}
\end{equation}
The condition $p(n^*) n^* \Delta n^*  = O(1)$
yields
\begin{equation}
 p(n^*)       \sim   L^{-3/2} \left[\ln L\right]^{1/2}\;,
\label{p2nstar}
\end{equation}
and the weight $w$ scales  as $1/L$.

The scaling  behaviour of the peak in $p(n)$ implies that  the
size of the condensates, (\ref{n2star}), is extensive.
Also, up to logarithmic corrections, the width of the peak (\ref{delnstar2})
scales as the square root of its position. Thus the peak is
sharp  as in the usual condensation scenario associated with
(\ref{ucond}).
The number of condensates is
\begin{equation}
q = \frac{\rho- \rho_c}{a} \;,
\end{equation}
where  following our previous discussion we consider
only densities $\rho$ for which $q$ is an integer.
We will comment on other densities in the next subsection.

To locate the position of the dip,  $n_{min}$, in $p(n)$,
we balance the first two terms in the extremum equation (\ref{phi2d0})
which yields
\begin{equation}
n_{min} \sim \frac{1}{h(L)} \sim \frac{L}{\ln L}
\label{n2min}
\end{equation}
and
\begin{equation}
p(n_{min}) \sim  \frac{1}{n_{min}^b} \sim  \left(\frac{\ln L}{L}\right)^b\;.
\label{p2nmin}
\end{equation}

It is interesting to note that an alternative way
of obtaining the scaling behaviour of the peak
of $p(n)$ in the  case (\ref{u2}) is by
taking the limit $k= aL$ in the results
(\ref{nstar},\ref{delnstar},\ref{pnstar})
and (\ref{nmin},\ref{pnmin}) for the algebraic case (\ref{u1}).

%%%%%%%%%%%%%%%%%%%%%%%%%%%%%%%%%%%
\subsection{Numerical results}
\label{numsim_Exp}
%%%%%%%%%%%%%%%%%%%%%%%%%%%%%%%%%%%

We start by considering a case with integer $(\rho- \rho_c)/a$.
The results of numerical simulations for a system with hopping rates
given by (\ref{u2}), size $L=1000$, $b=4$, $a=1$ and
$\rho=2.5>\rho_c=0.5$ are displayed in Figure~\ref{sim_exp_int}. The
probability distribution is given in Figure~\ref{sim_exp_int}(a) where
we can clearly see a narrow peak around $n \approx L=1000$.
The simulation results are given by the circles and are compared to
a numerical calculation of the exact canonical expression
(\ref{pnLN}) represented by the full line. Since the critical
density is 0.5 we expect the excess density $2.5-0.5 = 2$ to be
distributed among two condensates. The occupation of each condensate is expected to
be approximately equal to $L=1000$. A typical snapshot is
presented in Figure~\ref{sim_exp_int}(b) where the occupation of the
lattice is plotted.

%%%%%%%%
\begin{figure}[htb]
\begin{center}
\includegraphics[width=12cm]{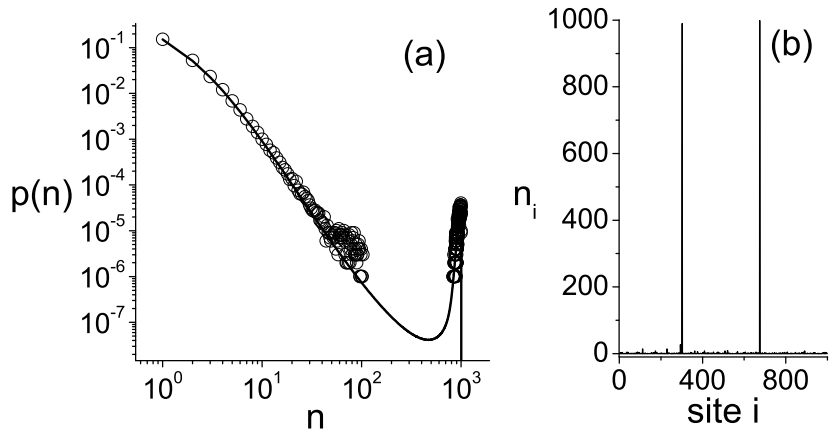}
\caption{\label{sim_exp_int} (a) A logarithmic plot of $p(n)$ 
in the case of algebraic hopping rates obtained
 from numerical simulation, represented by circles (o), compared to
 that calculated from the exact canonical expression (\ref{pnLN}),
 represented by the solid line.  (b) Site occupations in a snapshot of
 the system obtained from numerical simulation for $u(n)$ given by
 (\ref{u2}), $L=1000$, $\rho=2.5$, $b=4$ and $a=1$.

 %Fig 21 YS thesis
}
\end{center}
\end{figure}
%%%%%%%%

We now proceed to examine a case with non-integer $(\rho- \rho_c)/a$.
The results of numerical simulations for a system with hopping rates
given by (\ref{u2}), size $L=1000$, $b=4$, $a=1$ and
$\rho=2.3>\rho_c=0.5$ are displayed in
Figure~\ref{sim_exp_non_int}. The probability distribution is given in
Figure~\ref{sim_exp_non_int}(a) where we can clearly see a rather
broad peak around $n \approx L=1000$.  The simulation results are
given by the circles and are compared to a numerical calculation of
the exact canonical expression (\ref{pnLN}) represented by the full
line. Since the critical density is 0.5 we expect the excess density
$2.3-0.5 = 1.8$ to be distributed among two condensates. Even though
the occupation of each condensate fluctuates in time between $0.8$ and
$1$ they spend the majority of the time in either and not in
transition  as can be seen in Figure~\ref{cond_oscil}.  The resulting
distribution has two peaks, one centred around $n_1^*\approx 0.8L=800$
and the second around $n_2^*\approx L=1000$, which yield  the
appearance of a single  broader peak in Figure~\ref{sim_exp_non_int}(a). 
A typical snapshot is presented in
Figure~\ref{sim_exp_non_int}(b) where the occupation of the lattice is
plotted.

%%%%%%%%
\begin{figure}[htb]
\begin{center}
\includegraphics[width=12cm]{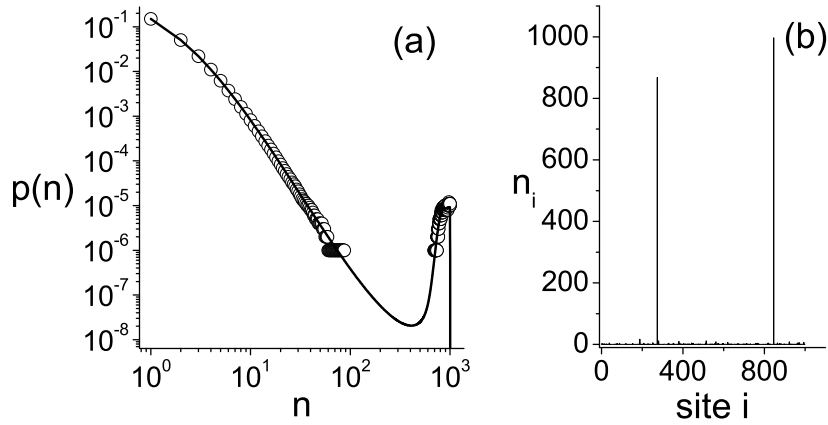}
\caption{\label{sim_exp_non_int} (a) A logarithmic plot of $p(n)$
in the case of algebraic hopping rates
 obtained from numerical simulation, represented by circles (o),
 compared to that calculated from the exact canonical expression
 (\ref{pnLN}), represented by the solid line .  (b) Site occupations
 in a snapshot of the system obtained from numerical simulation for
 $u(n)$ given by (\ref{u2}), $L=1000$, $\rho=2.3$, $b=4$ and $a=1$.

 %Fig 21 YS thesis
}
\end{center}
\end{figure}
%%%%%%%%

%%%%%%%%
\begin{figure}[htb]
\begin{center}
\includegraphics[width=12cm]{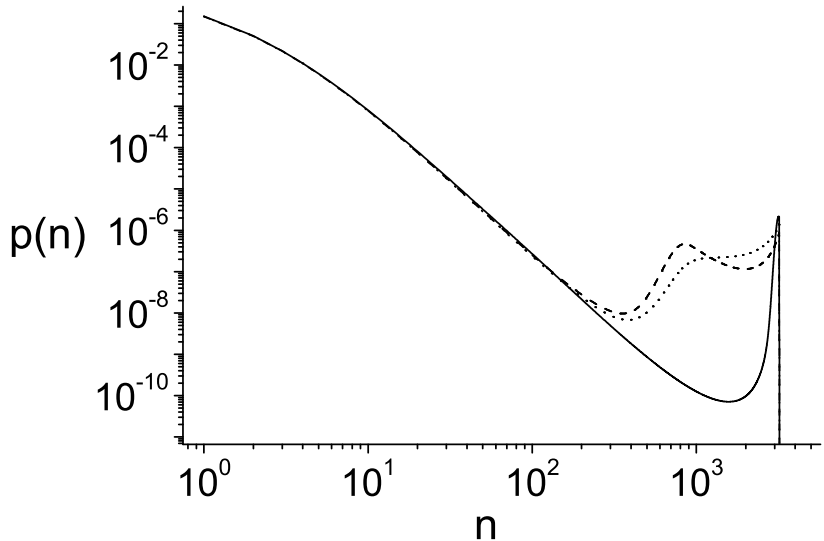}
\caption{\label{pnDiffR} The canonical probability
distribution calculated numerically using (\ref{pnLN}) for $L=3200$,  
in the case of exponential hopping rates with $b=4$
$a=1$ and $\rho = 1.5$, 1.75 and 2.75 for the full, dotted and dashed lines respectively.
}
\end{center}
\end{figure}
%%%%%%%%

In Figure~\ref{pnDiffR}, we compare the numerical solution for the
canonical distribution $p(n)$ for $\rho=1.5$ (integer $(\rho-
\rho_c)/a$) with that corresponding to $\rho=1.75$ and $\rho=2.75$
(where $(\rho- \rho_c)/a$ is non-integer)  for $L=3200$, $b=4$ and
$a=1$. For the above parameters with a density of $\rho=1.5$ there
is a single condensate corresponding to the single peak at
$aL=3200$. On the other hand for $\rho=2.75$ there are two
condensates corresponding densities $(\rho-\rho_c-a)L=800$ and
$aL=3200$, resulting in two distinct peaks in the distribution
function. This is verified in Figure~\ref{pnDiffR}.  Clearly, the
results for non-integer $(\rho- \rho_c)/a$ do not agree with the
grand canonical analysis in which there is a single peak
corresponding to the multiple condensates.

%%%%%%%%
\begin{figure}[htb]
\begin{center}
\includegraphics[width=15cm]{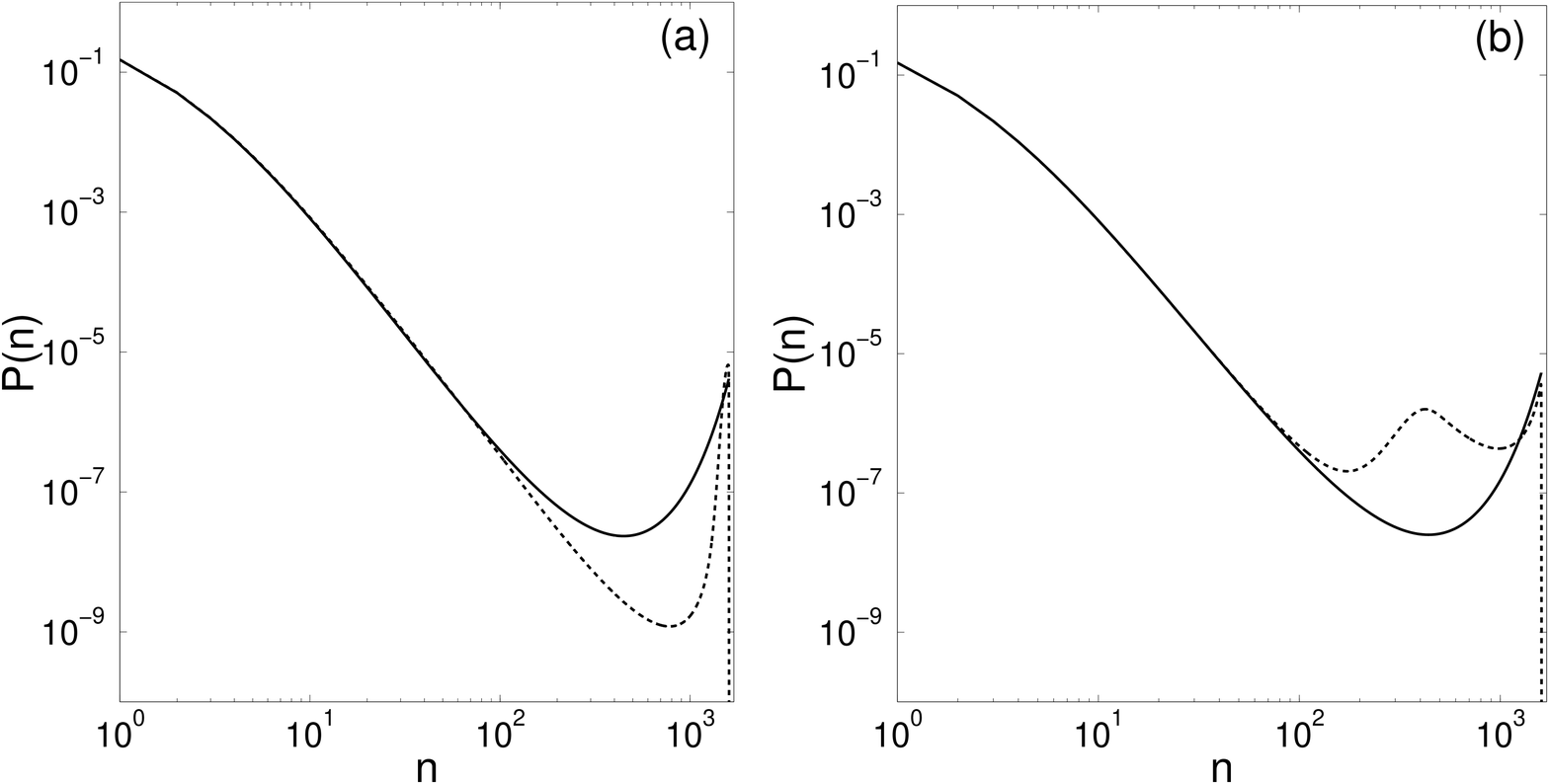}
\caption{\label{grand} A comparison of the canonical probability
(\ref{pnLN}) and the grand canonical probability (\ref{pn}),
 represented by the dashed and
full lines respectively.
The distributions were calculated numerically for  exponential hopping rates
with $L=1600$,  $b=4$
$a=1$ and $\rho = 1.5$  and $\rho = 1.75$ in figures (a) and (b) respectively. }
\end{center}
\end{figure}
%%%%%%%%

Since the steady state properties of the distribution given in
Section~\ref{secexp} were calculated in the grand canonical
ensemble, in which the distribution exhibits only a single peak, we
wish to compare the distribution obtained in the canonical ensemble
with the one obtained in the grand canonical ensemble.  In
Figure~\ref{grand} the probability distribution for the parameters
$L=1600$, $b=4$ and $a=1$ was calculated numerically both in the
grand-canonical ensemble (\ref{pn}) and in the canonical ensemble
(\ref{pnLN}). In Figure~\ref{grand}(a) the distributions are
compared for $\rho=1.5$, a density in which a single condensate
exits. In Figure~\ref{grand}(b) the distributions are compared for
$\rho=1.75$ a density in which two condensates exist. It can be seen
in Figure~\ref{grand}(a) and(b) that the distributions obtained form
the two ensembles are not equivalent; they differ mostly at large
$n$. On the other hand, for small $n$, namely in the fluid phase,
the distributions obtained from the two ensembles agree.

 For large occupations not only does the size dependent constraint on
 the number of particles affect the distribution but so does the size
 dependent term in the hopping rate (\ref{u2}).  This endogenous size
 dependence yields a different form for the distribution at large
 $n$. The height of the peak $p(n^*)$ is the same in both ensembles
 for the two cases in Figures~\ref{grand}(a) and (b).  However, the
 dip $p(n_{min})$ is different for the two ensembles even for the
 single condensate case in Figures~\ref{grand}(a).  The scaling
 behaviour in the canonical ensemble of the distribution as a function
 of the system size can be calculated numerically using (\ref{pnLN})
 and compared with the prediction in the grand-canonical ensemble
 (\ref{delnstar2},\ref{p2nstar},\ref{n2min},\ref{p2nmin}).

%%%%%%%%
\begin{figure}[htb]
\begin{center}
\includegraphics[width=10cm]{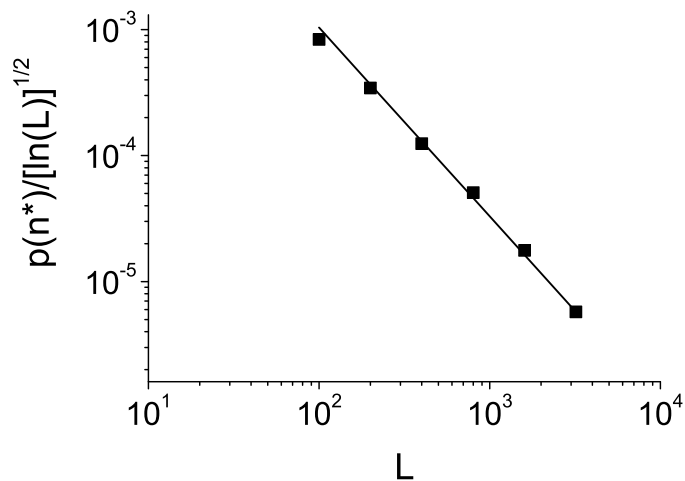}
\caption{\label{scale_pnstar}
The scaling of the condensate peak $p(n^*)/\left[\ln L \right]^{1/2}$ as a
function of the system size $L$ in the case of exponential hopping rates (\ref{u2})
with $b=4$, $\rho=2.5$ and $a=1$.
The probability $p(n^*)$ for each system size $L$ was calculated numerically
using (\ref{pnLN}).
The data is compared to a linear slope of $-3/2$.}
\end{center}
\end{figure}
%%%%%%%%

In Figure~\ref{scale_pnstar} we check  that in the case of integer
$(\rho-\rho_c)/a$ the grand canonical analysis yields the leading
scaling properties of the peak calculated  for the canonical
distribution in finite size systems. In the Figure, $p(n^*)/\left[
\ln L \right]^{1/2}$ is plotted against $L$. A  slope which agrees
with the expected value of $-3/2$ is seen, confirming the $L^{-3/2}$
dependence. However to test the logarithmic corrections more
detailed numerical data are required. 

For the non integer $(\rho-\rho_c)/a$ case the grand canonical analysis is not valid
and as a result, the scaling relations developed using the grand canonical analysis do not hold.
This was confirmed numerically using (\ref{pnLN}) to calculate the scaling in the canonical ensemble
 and comparing with the prediction in the grand-canonical ensemble
 (\ref{delnstar2},\ref{p2nstar},\ref{n2min},\ref{p2nmin}).

%%%%%%%%
\begin{figure}[htb]
\begin{center}
\includegraphics[width=15cm]{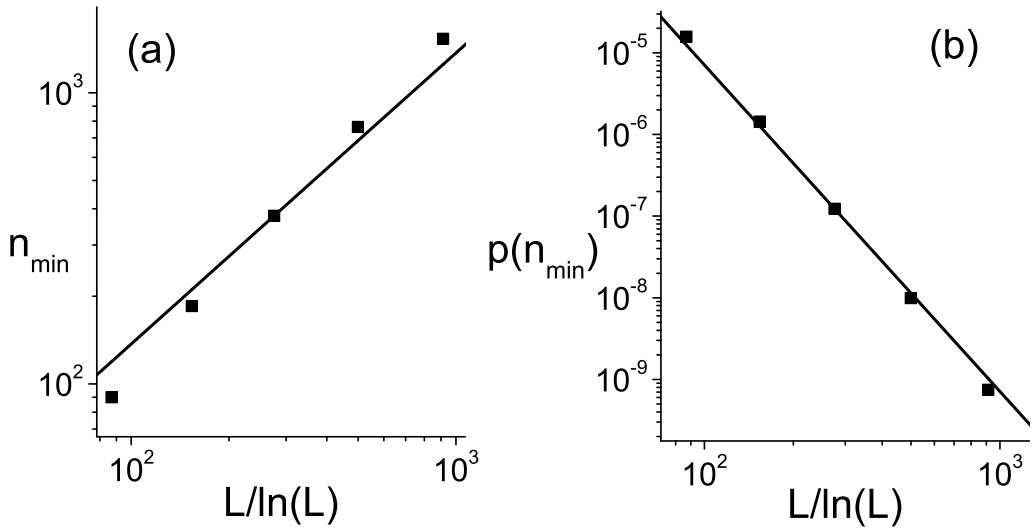}
\caption{\label{scale_nmin_u2} The scaling of the dip as a function of
the system size $L$ in the case of exponential hopping rates (\ref{u2})
with $b=4$, $\rho=2.5$ and $a=1$.  The probability
$p(n_{min})$ and location of the dip $n_{min}$ were calculated
numerically for each system size $L$ using (\ref{pnLN}).
 In (a) and (b) $n_{min}$ and $p(n_{min})$ are plotted
respectively as a a function of $L/\ln(L)$ and are compared to a
linear fit of the data points. The data points for the various system
sizes are compare to linear slopes of 1 and $-4$ in (a) and (b)
respectively.}
\end{center}
\end{figure}
%%%%%%%%

As noted above in the discussion of Figure~\ref{grand}, the
canonical and grand canonical distributions appear to differ in the
dip region. However it may be that scaling is still correctly
predicted by the grand canonical analysis.  In
Figure~\ref{scale_nmin_u2} we check that in the case of integer
$(\rho-\rho_c)/a$ the grand canonical analysis yields the leading
scaling properties of the dip calculated for the canonical
distribution in finite size systems. In Figure~\ref{scale_nmin_u2}(a)
the dip location $n_{min}$ is plotted against $L$.  A linear fit for
the data produced a slope of $1.2\pm0.2$. In the figure the data is
compared with a slope of $1$ corresponding to the scaling relation
(\ref{n2min}).  In Figure~\ref{scale_nmin_u2}(b) the dip height
$p(n_{min})$ is plotted against $L$.  A linear fit for the data
produced a slope of $-4.2\pm0.2$. In the figure the data is compared
with a slope of $-4$ corresponding to the scaling relation
(\ref{p2nmin}).  This is not the case for the non integer
$(\rho-\rho_c)/a$ since the dip is not well defined as the number of
humps may be larger than one.

We conclude that for the case of integer $(\rho-\rho_c)/a$ the
steady state properties of the distribution given in
Section~\ref{secexp} computed for the grand canonical ensemble are
able to describe the properties in the canonical ensemble. However,
it seems as if the analysis breaks down for the case of non integer
$(\rho-\rho_c)/a$ where the two descriptions are qualitatively
different in the description of the condensates.

%%%%%%%%%%%%%%%%%%%%%%%%%%%%%%%%%%%
\section{Dynamics}\label{dynamics}
%%%%%%%%%%%%%%%%%%%%%%%%%%%%%%%%%%%

\subsection{Condensate creation and evaporation timescales}
We are interested in the
dynamics of condensate formation
for systems described by the hopping rates (\ref{u1},\ref{u2}).
Previous studies of condensate  formation and dynamics for the case
(\ref{ucond}) have been carried out \cite{GSS03,Godreche03,GL05}.
We identify two timescales
associated with  creation and evaporation of  (meso)condensates in
the system. To identify these timescales we note that
the size of the
(meso)condensates fluctuates around $n^*$. These
fluctuations are occasionally  large enough to cause a
condensate to evaporate.
The typical time  for a condensate to exist
at a given site
before it evaporates  is termed the evaporation
time and is denoted by $\tau_e$. Once a condensate has evaporated the
particles are redistributed among the other sites and this can
result in formation of another condensate. The
typical time that  a given site exists in the fluid phase
before a condensate is created at that site is
termed the creation time and is denoted by $\tau_c$.
%%%%%%%%
\begin{figure}[htb]
\begin{center}
\includegraphics[width=10cm]{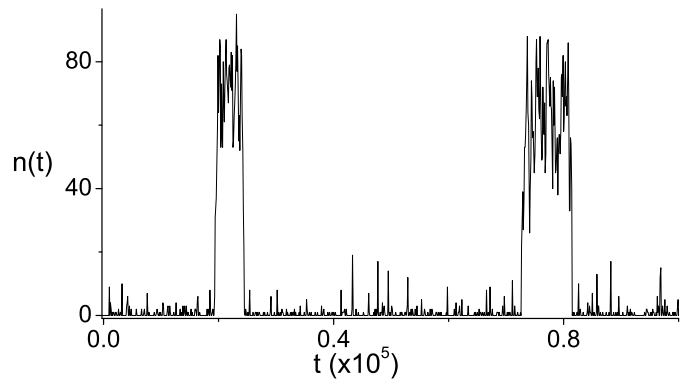}
\caption{\label{tevo} The time series for an occupation of a single
site for a lattice of size $L=100$ in the case of algebraic hopping
rates (\ref{u1}) with $b=4$, $\rho=4$, $c=1$ and $k=10$.  }
\end{center}
\end{figure}
%%%%%%%%

In Figure~\ref{tevo} we display the time series
of the  occupation number of a
typical site, for the algebraic case (\ref{u1}). One clearly sees sharp
transitions between the fluid state and condensed state. By averaging
over a long time series the evaporation and creation times may be evaluated.
The results are illustrated in
Figure~\ref{tscale}.  Similar  plots for the exponential case
are illustrated in Figure~\ref{tscale2} for an integer number of condensates.

%%%%%%%%
\begin{figure}%[htb]
\begin{center}
\includegraphics[width=10cm]{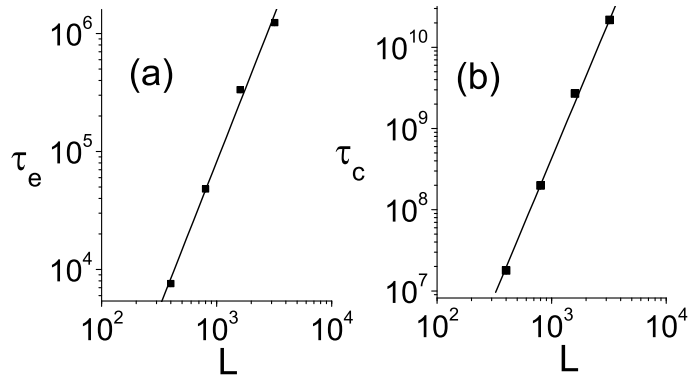}
\caption{\label{tscale} The evaporation time $\tau_e$ and creation
time $\tau_c$ as a function of the lattice size $L$ for an algebraic hopping rate (\ref{u1})
 calculated numerically from simulations. The result is
plotted for algebraic hopping rates with $b=4$, $\rho=3$ and $k=5$. The
results were averaged over 100 repetitions for $\tau_e$ and $\tau_c$
in figures (a) and (b) respectively. The results are compared with a
linear slope $(b-1)k/(k+1)=2.5$ in figure (a) and $bk/(k+1)=3.37$ in
figure (b).  }
\end{center}
\end{figure}
%%%%%%%%

In order to estimate
the evaporation
time $\tau_e$ and creation time $\tau_c$ we use the Arrhenius law.
As we have seen, the probability distribution of the occupation
of a given site is described by an effective potential $\Phi(n)$
as illustrated in Figure~\ref{phi}.
In the condensed phase $\Phi(n)$
exhibits two valleys. The left valley corresponds to fluid states and
the right valley centred around $n^*$ corresponds to
condensate states. The potential barrier centred around $n_{min}$
corresponds to the dip region in the probability distribution.
Using the Arrhenius law, a condensate will form on the site
once the occupation of the site crosses the
barrier from the left to the right valley. The timescale for
creation of a condensate is then proportional to the ratio of the
probability of being in the fluid to the probability of being at the
creation threshold $n_{min}$. Thus, since the probability of being in
the fluid is of the order of $1$, we obtain
\begin{equation}
\tau_c\,\sim\,\frac{1}{p(n_{min})}\;.
\label{tauc}
\end{equation}

For the evaporation time of a condensate, the timescale is
proportional to the ratio of the  probability for a condensate, which is of the
order of $w$, to the probability of being at the evaporation threshold which is again at
$n_{min}$,
\begin{equation}
\tau_e\,\sim\,\frac{w}{p(n_{min})}\;.
\label{taue}
\end{equation}
Note  that on general grounds the ratio of the two timescales has to satisfy
\begin{equation}
\frac{\tau_e}{\tau_c}\,\sim\,\frac{w }{1-w}
\label{taugen}
\end{equation}
where $w/(1-w)$ is the ratio of the probabilities of a site being in
a condensate or a fluid  state.
This follows from a steady-state  condition that
the rate per site  of condensate creation  $(1-w)/\tau_c$
is balanced by the rate per site of condensate evaporation $w/\tau_e$.
For small $w$, (\ref{taue}) and (\ref{tauc}) satisfy (\ref{taugen}).

%%%new figure
\begin{figure}[htb]
\begin{center}
\includegraphics[width=10cm]{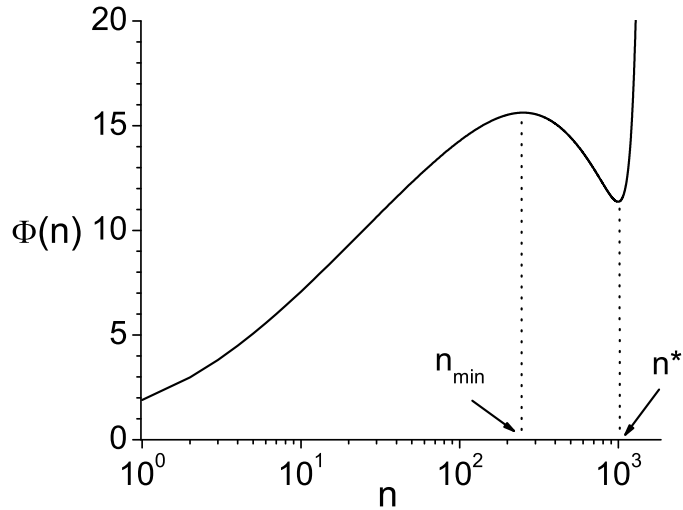}
\caption{\label{phi} The effective potential $\Phi(n)$ for a lattice
of size $L=1600$ in the case of algebraic hopping rates
with $b=4$, $\rho=4$, $c=1$ and $k=10$. $\Phi(n)$
was calculated numerically  from the exact expression (\ref{pnLN})
for the
 hopping rate $u(n)$ given
by  (\ref{u1}) .  }
\end{center}
\end{figure}
%%%%%%%%

%%%%%%%%%%%%%%%%%%%%%%%%%%%%%%
\subsection{Numerical results for algebraically increasing case}
%%%%%%%%%%%%%%%%%%%%%%%%%%%%%%

For the algebraic hopping rate (\ref{u1}), the creation and
evaporation times scale with the system size as
\begin{eqnarray}\label{algebraic_t}
 \tau_c  &\sim& {\bigg(\frac{L}{\ln L }\bigg)}^{bk/(k+1)} \quad , \quad
 \tau_e  \sim  L^{(b-1)k/(k+1)}\, \left[\ln L\right]^{-(1+bk)/(k+1)}\;.
 \end{eqnarray}
 Using numerical simulations for for $b=4$, $\rho=3$ and $k=5$ the
evaporation time $\tau_e$ and creation time $\tau_c$ are plotted in
Figure~\ref{tscale} as a function of the lattice size $L$ . In
Figure~\ref{tscale}(a) the evaporation time $\tau_e$ as a function of
lattice size is compared on a double logarithmic scale to a linear
slope of $(b-1)k/(k+1)=2.5$.  In Figure~\ref{tscale}(b) the creation
time $\tau_c$ as a function of lattice size $L$ is compared on a
double logarithmic scale to a linear slope of $bk/(k+1)=3.37$.

%%%%%%%%%%%%%%%%%%%%%%%%%%%%%%
\subsection{Numerical results for exponentially increasing case}
%%%%%%%%%%%%%%%%%%%%%%%%%%%%%%
In the following, we separate the discussion of the case of an integer $(\rho-\rho_c)/a$
from that of a non integer $(\rho-\rho_c)/a$.
To carry out an analogous analysis for the exponential hopping rate (\ref{u2})
with an integer $(\rho-\rho_c)/a$.
we require an expression for how $n_{min}$ scales with $L$.
As we have discussed, the hopping rate (\ref{u2})
is the same as the usual case (\ref{ucond}) but with the introduction  of a
sharp cut-off at $n=aL$. Thus  we expect $p(n)$ to remain the
same as that for (\ref{ucond}) up to the condensate peak
which takes place at $n= O(L)$. Therefore $n_{min}= O(L)$,
as in the usual case (\ref{ucond}),
and $p(n_{min})\sim 1/L^b$.
This estimate inserted into
(\ref{tauc},\ref{taue}) yields the following leading behaviour
for the creation and evaporation timescales in the exponential hopping rate
case
\begin{eqnarray}\label{exp_t}
 \tau_c &\sim& L^b \quad ,
\quad  \tau_e \sim L^{b-1}\;.
\end{eqnarray}

%%%new figure
\begin{figure}[htb]
\begin{center}
\includegraphics[width=10cm]{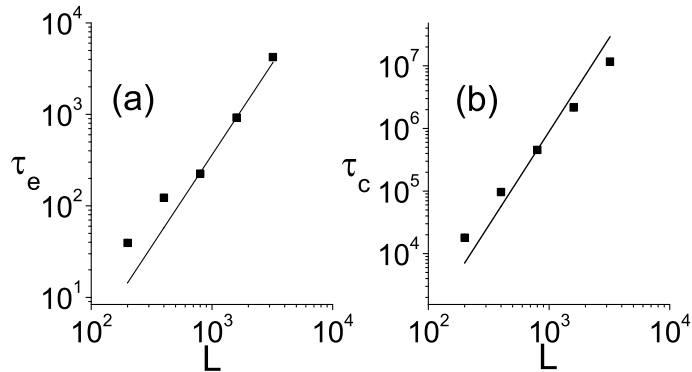}
\caption{\label{tscale2}
The evaporation time $\tau_e$ and creation time $\tau_c$
 as a function of the lattice
size $L$ for an exponential hopping rate (\ref{u2}) were calculated numerically from
simulations. 
The  results are  plotted for $b=3$, $\rho=4$ and $a=1$.
 The results where averaged over 100 repetitions for $\tau_e$
and $\tau_c$ in figures (a) and (b) respectively.
 The results are compared to a
linear  slope $b-1=2$ in figure (a) and
$b=3$ in figure (b) . }
\end{center}
\end{figure}
%%%%%%%%
We test the leading $L$ behaviour using numerical simulations for
$b=3$, $\rho=4$ and $a=1$, which corresponds to an integer
$(\rho-\rho_c)/a=3$, the evaporation time $\tau_e$ and creation time
$\tau_c$ are plotted in Figure~\ref{tscale2} as a function of the
lattice size $L$ . In Figure~\ref{tscale2}(a) the evaporation time
$\tau_e$ as a function of lattice size is compared on a double
logarithmic scale to a linear slope of $(b-1)=2$.  In
Figure~\ref{tscale2}(b) the creation time $\tau_c$ as a function of
lattice size $L$ is compared on a double logarithmic to a linear slope
of $b=3$.  These results are consistent with the leading $L$ behaviour derived for the grand canonical ensemble (\ref{exp_t}), although more extensive simulations would be desirable.

On the other hand, for the case of non-integer $(\rho-\rho_c)/a$ the
the leading $L$ behaviour of the creation time $\tau_c$ and of the
evaporation time $\tau_e$ which were calculated within the grand
canonical ensemble are invalid.  The Arrhenius law approach is based
on a distribution with a single peak and a single dip which is not
the case for the non-integer $(\rho-\rho_c)/a$ as was shown above.
Thus, for the non-integer $(\rho-\rho_c)/a$ not only is the
distribution obtained within the grand canonical ensemble not valid,
so is the framework in which we calculate the evaporation and
creation times.

%%%%%%%%%%%%%%%%%%%%%%%%%%%%%%%%%%%
\section{Conclusions}\label{conc}
%%%%%%%%%%%%%%%%%%%%%%%%%%%%%%%%%%%
In this work we have shown how the introduction of non-monotonic
hopping rates $u(n)$ into the ZRP can produce a condensation
transition into many condensates. The algebraic choice of hopping rate
(\ref{u1}) results in mesocondensates whose number grows
subextensively with system size $L$.  
On the other hand, the
exponential choice of hopping rate (\ref{u2}) can result in a finite
number of extensive condensates.

Related condensation phenomena, which result in a large number of
mesocondensates, have previously discussed in the context of
non-conserving ZRP\cite{AELM07}.  Thus our results  imply that the
steady states of such models can  be described
by a conserving ZRP with an  effective hopping rate which is non-monotonic.

We have analysed the single-site distribution $p(n)$ within the grand
canonical ensemble and in particular analysed the scaling behaviour of
the condensate peak. We have computed the scaling behaviour of the
position, height, width of the peak and also the position and
height of the dip
between the condensate peak and the part of the distribution
representing the fluid as illustrated in Figure~\ref{pdetail}.

In the case of algebraically increasing hopping rates (\ref{u1}) the
predictions of the grand canonical analysis are well borne out by
numerical computation within the canonical ensemble.  Ignoring
logarithmic corrections the scaling of the condensate peak is as follows:
the position of the peak scales as $n^* \sim
L^{k/(k+1)}$; the width scales as $\Delta n^* \sim L^{k/(k+1)}$; the
height of the peak scales as $p(n^*) \sim L^{-2k/(k+1)}$, and the
weight of the peak scales as $w \sim L^{-k/(k+1)}$.  The behaviour of
$w$ implies that the number of mesocondensates scales as
$L^{1/(k+1)}$.  The peak is not sharp since $\Delta n^*/ n^* \to 0 $
only as a power of $\ln L$. Thus we term it a `weak peak'.  The weak
peak is similar to that determined in the analysis of a non-conserving
ZRP \cite{AELM07}.

For the exponential case (\ref{u2}) the situation is more subtle.  For
the case of integer values of $(\rho-\rho_c)/a$, our numerical results
suggest that the scaling predicted by the grand canonical analysis is
correct even though the grand canonical and canonical distributions do
not appear to coincide.  The scaling of the condensate peak, excluding
any logarithmic corrections, is as follows: the position of the peak scales as
$n^* \sim L$; the width scales as $\Delta n^* \sim L^{1/2}$; the
height of the peak scales as $p(n^*) \sim L^{-3/2}$, and the weight of
the peak scales as $w \sim L^{-1}$.  The behaviour of $w$ implies that
the number of mesocondensates is finite.  In the exponential case the
peak is `sharp' since $\Delta n^*/ n^* \to 0 $ as a power of $L$.

For the case of non-integer values of $(\rho-\rho_c)/a$, on the other hand,
it appears that the results of the  grand canonical analysis break down.

It is of interest to compare the scaling form we have determined of
the peak in $p(n)$ corresponding to (meso)condensates to the usual
scenario of condensation into a single site exhibited, for example, by
the hopping rate (\ref{ucond}).  For the single condensate case, in
\cite{MEZ05,EMZ06} the condensate peak, denoted $p_{cond}(n)$ in that
work, has been computed.  The scaling form of the peak is as follows:
the position scales as $n^* \sim L$; the width scales as $\Delta n^*
\sim L^{1/2}$ for $b>3$ and $\Delta n^* \sim L^{1/(b-1)}$ for $3>b>2$;
the height scales as $p(n^*) \sim L^{-3/2}$ for $b>3$ and $p( n^* )
\sim L^{-b/(b-1)}$ for $3>b>2$; and the weight of the peak scales as
$w \sim L^{-1}$.  The peak is sharp, as in the case we have studied of
a finite number of condensates.  However, in the case of a finite
number of condensates, our grand canonical predictions for
the scaling of the width and height of the peak do not depend on $b$.

We note that for the usual single condensate scenario the results
quoted above for $p_{cond}(n)$ were computed within the canonical
ensemble \cite{MEZ05,EMZ06}.  It would be of interest to see if our
results can be recovered in a canonical calculation.  Also, an
analysis within the canonical ensemble is required to obtain the
scaling form the condensate peak in the exponential case when
$(\rho-\rho_c)/a$ is non-integer.

In studying the dynamics of the condensates we identified two
timescales, one for the creation of a condensate at a given site and
the other for the evaporation of a condensate at a given site.  The
scaling of these timescales with the system size has been studied
within the phenomenological Arrhenius approach; how these timescales
scale with system size is determined by the scaling of height of the
dip and the weight of the condensate peak.  We have found that the
predictions compare well with numerical simulations in the algebraic
case.  In the exponential case the numerical results are consistent with
the Arrhenius predictions, but  more extensive simulations would be
desirable.

%%%%%%%%%%%%%%%%%%%%%%%%%%%%%%%%%%%
%%%%%%%%%%%%%%%%%%%%%%%%%%%%%%%%%%%
\ack
We thank Attila Rakos for useful discussions.
This study was partially supported by the Israel Science Foundation (ISF).
Visits of MRE to the Weizmann Institute were supported by the
Albert Einstein Minerva Center for Theoretical Physics. Visits
of DM to Edinburgh were supported by the EPSRC programme grant GR/S10377/01.
We thank the Isaac Newton Institute in Cambridge, UK for kind hospitality during the programme `Principles of Dynamics of Nonequilibrium Systems'
where part of this project was carried out.

\end{document}